\newcommand{\hw}{\ensuremath{\hbar\Omega}}
\newcommand{\outerprod}[2]{\vert #1 \rangle\!\langle #2 \vert}
\newcommand{\ket}[1]{|#1\rangle}

\documentclass[conference]{IEEEtran}
\IEEEoverridecommandlockouts
\usepackage{cite}
\usepackage[colorlinks=true, allcolors=blue]{hyperref}
\usepackage{amsmath,amssymb,amsfonts}
\usepackage{algorithmic}
\usepackage{graphicx}
\usepackage{textcomp}
\usepackage{xcolor}
\def\BibTeX{{\rm B\kern-.05em{\sc i\kern-.025em b}\kern-.08em
    T\kern-.1667em\lower.7ex\hbox{E}\kern-.125emX}}
\begin{document}

\title{Efficacious qubit mappings for quantum simulations of the $^{12}$C rotational band\\
\thanks{This work was supported by the U.S. Department of Energy
(DE-SC0023694). SR acknowledges support from the School of Electrical and Computer Engineering
at Cornell University, from the U.S. National Science Foundation under Grant No. 2315398, and from AFRL under agreement no. FA8750-23-2-0031. We acknowledge discussions with the Quantum Science Center at ORNL. This work benefited from high performance computational resources provided by LSU (www.hpc.lsu.edu), the National Energy
Research Scientific Computing Center (NERSC), a U.S.
Department of Energy Office of Science User Facility at
Lawrence Berkeley National Laboratory operated under
Contract No. DE-AC02-05CH11231, as well as the Frontera
computing project at the Texas Advanced Computing
Center, made possible by National Science Foundation
award OAC-1818253.}
}

\author{\IEEEauthorblockN{Darin C. Mumma\IEEEauthorrefmark{1}, Zhonghao Sun\IEEEauthorrefmark{2},  Alexis Mercenne\IEEEauthorrefmark{1}, 
Kristina D. Launey\IEEEauthorrefmark{1},\\
Soorya Rethinasamy\IEEEauthorrefmark{3}\IEEEauthorrefmark{4}\IEEEauthorrefmark{1} and James A. Sauls\IEEEauthorrefmark{1}\IEEEauthorrefmark{4}
} 

\IEEEauthorblockA{
\IEEEauthorrefmark{1}Department of Physics and Astronomy\\ Louisiana State University, Baton Rouge, LA 70803, USA
}

\IEEEauthorblockA{
\IEEEauthorrefmark{2}Quantum Science Center,\\ Oak Ridge National Laboratory, Oak Ridge, TN 37830, USA
 }

\IEEEauthorblockA{
\IEEEauthorrefmark{3}School of Applied and Engineering Physics,\\ Cornell University, Ithaca, New York 14850, USA
}

\IEEEauthorblockA{
\IEEEauthorrefmark{4}Hearne Institute for Theoretical Physics and Center for Computation and Technology,\\ Louisiana State University, Baton Rouge, Louisiana 70803, USA
}

 }

\maketitle

\begin{abstract}
Solving atomic nuclei from first principles places enormous demands on computational resources, which grow exponentially with increasing number of particles and the size of the space they occupy. We present first quantum simulations based on the variational quantum eigensolver for the low-lying structure of the $^{12}$C nucleus that provide acceptable bound-state energies even in the presence of noise. We achieve this by taking advantage of two critical developments. First, we utilize an almost perfect symmetry of atomic nuclei that, in a complete symmetry-adapted basis, drastically reduces the size of the model space. Second, we use the efficacious Gray encoding, for which it has been recently shown that it is resource efficient, especially when coupled with a near band-diagonal structure of the nuclear Hamiltonian.  
\end{abstract}


\section{Introduction}
Numerical simulations of nuclear structure and reactions are at the frontier of nuclear science research. However, the computational resources typically required scale exponentially with the number of particles and the size of the space they occupy. Quantum computing has the potential to revolutionize the way we compute through the superposition of qubit states\cite{Arute2019}. Earlier studies that aim to solve quantum many-body problems using noisy intermediate-scale quantum (NISQ) computers have been based on various approaches, including the variational quantum eigensolver (VQE)\cite{Peruzzo2014,Grimsley2019}, quantum phase estimation\cite{AspuruGuzik2005,Whitfield2011}, and quantum
imaginary time evolution\cite{Motta2019}. Among the various quantum algorithms, the VQE approach has been used to solve simplified nuclear science problems\cite{eugen2018} and more realistic problems\cite{PrezObiol2023} for its relatively shallower circuit depths and suitability for near-term quantum computers. Meanwhile, the development of efficient and robust quantum algorithms to solve the nuclear science problem in the Noise Intermediate-Scale Quantum
(NISQ) device era\cite{Preskill2018} could also help us to understand the entanglement of atomic nuclei, and foster algorithms for classical computing. 
\begin{figure*}[th]
\centering
\hspace{-0.5cm}
\includegraphics[width=.9\linewidth]{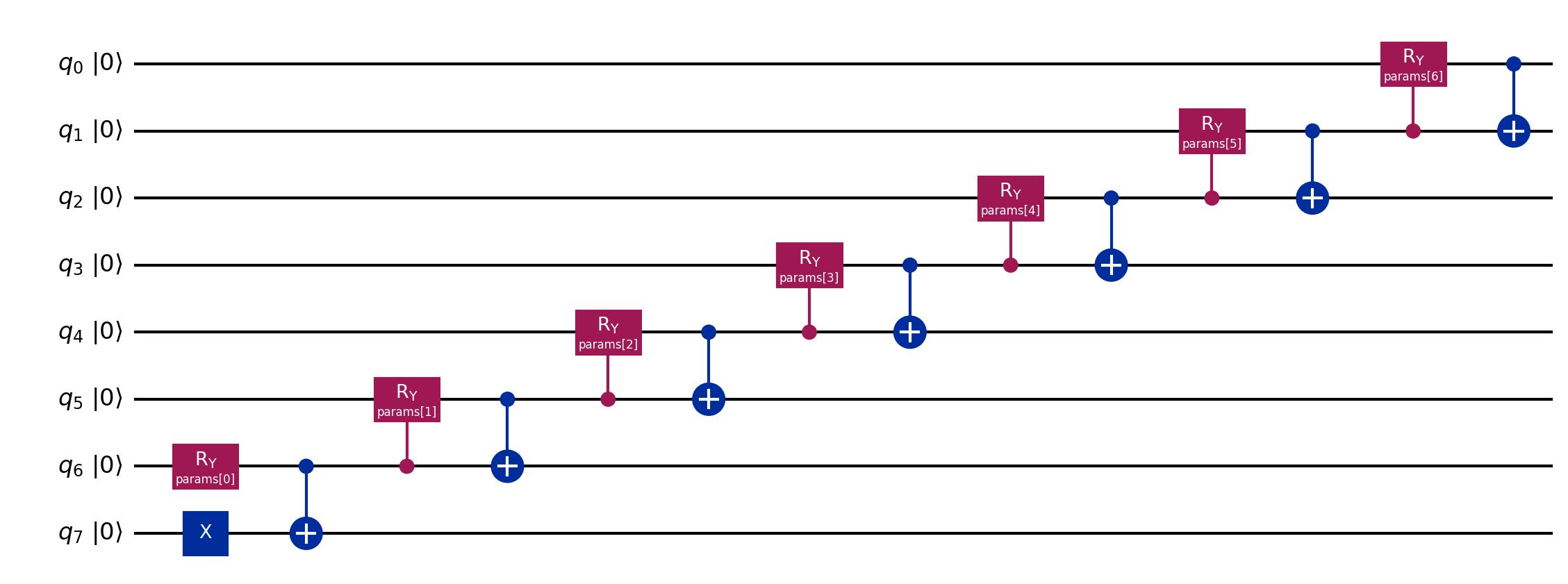}
\vspace*{-0.2cm}
\caption{\label{fig:Gray circuit}Recursive circuit ansatz to generate the superposition of the one-hot basis states used in Fig. (\ref{fig:grayvsOH}). The input to the circuit is $\ket{0}^{\bigotimes N}$, and $\text{params}[i]$ denote the parameters $\theta_{i}$ with the $R_{y}(2\theta_{i}) = e^{-i\theta_{i}Y_{i}}$ rotation gate.}
\end{figure*}

\begin{figure*}
\centering
\hspace{-1.1cm}
\includegraphics[width=0.8\linewidth]{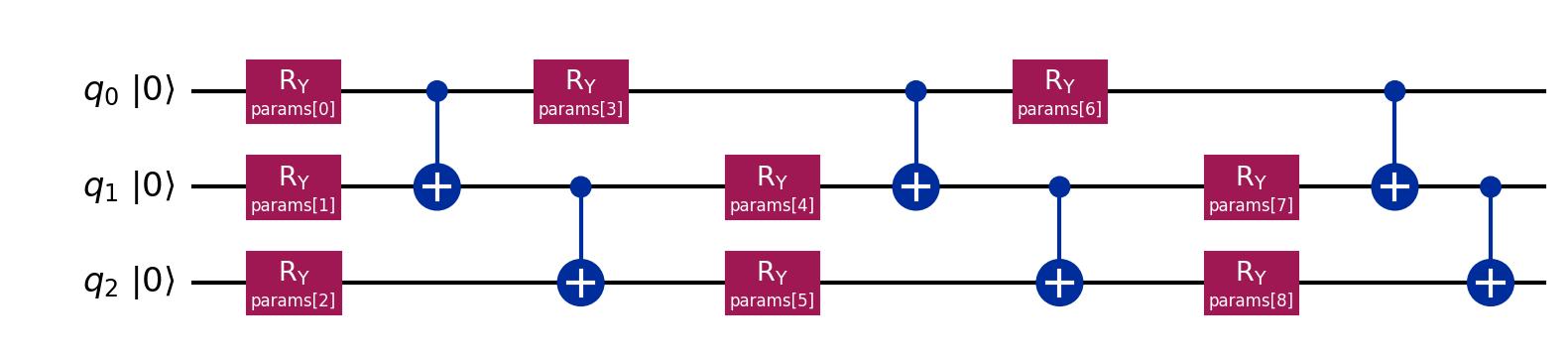}
\vspace*{-0.2cm}
\caption{\label{fig:OH circuit} Three-layered circuit ansatz to generate the superposition of the Gray basis states used in Fig. (\ref{fig:grayvsOH}). The input to the circuit is $\ket{0}^{\bigotimes N}$, and $\text{params}[i]$ denote the parameters $\theta_{i}$ with the $R_{y}(2\theta_{i}) = e^{-i\theta_{i}Y_{i}}$ rotation gate.}
\end{figure*}

Most nuclei are deformed, and one of the most challenging problems is explaining how fundamental interactions drive the deformation of atomic nuclei. The focus of this research is the structure of $^{12}$C. We perform quantum simulations for the ground state ($0^+_{\rm g.s.}$) and the first $2^+$ excited state of $^{12}$C, which are members of the ground-state rotational band and reflect strong collective correlations (e.g., see \cite{DytrychSBDV_PRL07}). Indeed, when starting with single-particle basis states of the harmonic oscillator (HO), the size of the model space required to describe the ground state, e.g., in fourteen HO shells is  $\sim 10^{11}$, which is beyond the memory capabilities of the best of today's petascale classical computers. Fortunately, when we use a complete symmetry-adapted correlated basis that recognizes an almost perfect symmetry of atomic nuclei~\cite{DytrychLDRWRBB20,LauneyMD_ARNPS21}, the symplectic Sp(3,R) symmetry \cite{Rowe85}, solutions can be achieved in model spaces that are orders of magnitude smaller. As a proof-of-principle, in this study, we provide solutions by adopting the nuclear Hamiltonian of \cite{DreyfussLTDB13} inspired by a (symplectic) symmetry effective field theory that captures most of the physics, namely, in excess of 85\% for each wavefunction (see the contribution of the dominant symmetry-adapted configuration in $0^+_{\rm g.s.}$ and $2^+$ of $^{12}$C from chiral potentials in ~\cite{Becker_BG2024}). This Hamiltonian has been shown to successfully describe the elusive Hoyle state in $^{12}$C~\cite{DreyfussLTDB13}. Remarkably, it further reduces the number of basis states to eight states only, enabling early experiments with twelve particles on NISQ processors. Following the first quantum applications to deuteron \cite{eugen2018}, a two-particle system, our quantum simulations use VQE to compute the energy of the twelve-particle nucleus. We use two different qubit encoding schemes, including one-hot and Gray encodings (see, e.g.,~\cite{Sawaya20}). We explore a recently developed algorithm that can efficiently solve any band-diagonal to full Hamiltonian matrices~\cite{rethinasamy2024neutronnucleus}. 
This approach can accommodate realistic nuclear potentials derived from chiral effective field theory, a key ingredient of long-term quantum simulations for \textit{ab initio} modeling of classically unsolvable atomic nuclei.

In particular, we investigate simulations of the $^{12}$C structure with shot noise and noise models from existing quantum devices, and explore the noise-resilient training method~\cite{Sharma_2020}. Our simulations reproduce the theoretical value of the $2^+$ excitation energy in $^{12}$C, even for noisy simulations suitable for the NISQ regime. We study the effectiveness of the one-hot and Gray encodings. We find that the Gray encoding allows an efficient scaling of the Hamiltonian complexity, by using only $\log_2(N)$ qubits for $N$ number of basis states, while achieving acceptable outcomes. Indeed, in ~\cite{rethinasamy2024neutronnucleus}, it has been shown that the Gray encoding is more resource efficient not only for tridiagonal Hamiltonians, as suggested earlier in ~\cite{Sawaya20}, but also for band-diagonal Hamiltonians with bandwidth up to $N$. 

\section{Nuclear Hamiltonian and quantum algorithm}
The nuclear Hamiltonian suitable for NISQ processors emerges from a symmetry effective field theory and consists of a fully microscopic nucleon-nucleon interaction \cite{DreyfussLTDB13}:
\begin{eqnarray}
    \label{effH0} 
   H_{\rm SymEFT} =&&\hspace{-0.6cm} H_0 + \chi \frac{\left( e^{-\gamma Q\cdot Q} -1 \right)}{2\gamma} \cr \approx&&\hspace{-0.6cm}  H_0  -\frac{\chi}{2} Q \cdot Q \sum_{k=0}^{\lceil K/2 \rceil -1} \frac{(-\gamma )^{k}(Q \cdot Q)^{k}}{(k+1)!},
\end{eqnarray}
where $H_0$ is the HO Hamiltonian, $Q
=\sum_i \sqrt{16\pi/5} r_i^2Y_{2}(\hat {\mathbf r}_i)$ is the mass quadrupole moment, and  ${\mathbf r}_i$ is the $i$-th particle coordinate. $K < N$ denotes the number of off-diagonals kept in calculations, with $2K+1$ giving the Hamiltonian matrix bandwidth, and $\chi$ and $\gamma$ are parameters of the nuclear potential. The corresponding calculations on classical computers have been earlier used to provide the first no-core shell-model description of the elusive Hoyle state in $^{12}$C that admits cluster substructures \cite{LauneyDD16,DreyfussLTDB13}. 
Our simulations start from  \eqref{effH0} with \hw \hspace{0.15cm}= 18 MeV, $\gamma=1.7 \times 10^{-4}$, and $\chi$ is defined in  \cite{DreyfussLTDB13}, in a model space of 14  HO shells (equivalently, maximum total HO excitations $N_{\rm max}=12$). The use of the symmetry-adapted basis enables applications of the quantum algorithm, discussed next, to eight non-negligible basis states.

We employ two different encoding schemes: the one-hot encoding and Gray encoding. Suppose the truncated Hilbert space is spanned by $N$ many-body basis states $(\ket{\psi_0}, \ket{\psi_1}, \cdots, \ket{\psi_{N-1}})$. The one-hot encoding maps these $N$ many-body basis states to $N$ qubit states: $ |\psi_i\rangle \to |0\cdots, 1_{i},0\cdots \rangle$. In contrast, the Gray encoding maps $N$ many-body basis states to $n=\lceil \log_2(N) \rceil$ qubits, and, e.g., requires 3 qubits for $N=8$:
\begin{center}
\begin{tabular}{ c c c }
    $\ket{\psi_0}$ & $\to$ & $\ket{000}$, \\ 
    $\ket{\psi_1}$ & $\to$ & $\ket{100}$, \\ 
    $\ket{\psi_2}$ & $\to$ & $\ket{110}$, \\ 
    $\ket{\psi_3}$ & $\to$ & $\ket{010}$, \\ 
    $\ket{\psi_4}$ & $\to$ & $\ket{011}$, \\ 
    $\ket{\psi_5}$ & $\to$ & $\ket{111}$, \\ 
    $\ket{\psi_6}$ & $\to$ & $\ket{101}$, \\ 
    $\ket{\psi_7}$ & $\to$ & $\ket{001}$. \\ 
\end{tabular}
\end{center}
The ordering of the different bitstrings of Gray encoding is not unique, but requires each bitstring to differ from its neighbor by a single bit, which, in turn, reduces the number of unitary operations in the quantum circuits. The particular form of the Gray encoding used in this work is the binary reflective Gray code, the same one used in \cite{rethinasamy2024neutronnucleus}.

Clearly, for the same $N$, the one-hot encoding needs exponentially more qubits. Hence, for $N \gtrsim 20$, simulations using the one-hot encoding become very challenging on classical architectures. For example, $N=32$ basis states requires 32 qubits and 61 two-qubit gates for the one-hot encoding, but only 5 qubits and 12 two-qubit gates for the Gray encoding with 3 layers \cite{rethinasamy2024neutronnucleus}.  In addition, the Gray encoding can utilize more parameters, which has been experimentally observed to help escape local minima. The reason is that these parameters provide additional degrees of freedom that the optimizer can utilize. Most importantly, errors in the one-hot encoding (e.g., erroneous flipping of a qubit) force the simulations to leave the model space of allowed states, making such simulations invalid. In contrast, the Gray encoding uses the entire model space (within a finite size), thereby, easily mitigating such errors. 

Following ~\cite{rethinasamy2024neutronnucleus}, the one-hot encoding maps a general Hamiltonian $H_{N, K}$ of size $N$ and bandwidth $2K+1$ as:
\begin{eqnarray}
    \hspace{-0.2cm} H_{N, K} \hspace{-0.05cm}=\hspace{-0.6cm}&& \frac{1}{2} \sum\limits_{m=0}^{N-1} \langle m \vert H \vert m \rangle (I_m - Z_m) +\cr
    && \sum\limits_{m=0}^{N-1} \sum\limits_{k=1}^{K} 
    \frac{\langle m+k \vert H \vert m \rangle  }{2} (X_m X_{m+k} + Y_m Y_{m+k}),
        \label{eq:HN_OneHot}
\end{eqnarray}
where $\langle m' \vert H \vert m \rangle$ are known matrix elements given by the Hamiltonian of  \eqref{effH0}, and $X$, $Y$, and $Z$ are Pauli matrices.

The Gray encoding maps the Hamiltonian $H_{N, K}$ as:
\begin{eqnarray}
\setlength\abovedisplayskip{0pt}
\label{eq:HN_Gray}
   H_{N, K} =&&\hspace{-0.6cm} \sum\limits_{m=0}^{N-1} \langle m \vert H \vert m \rangle G^0_m + \cr
    && \sum\limits_{m=0}^{N-1} \sum\limits_{k=1}^{K} \langle m+k \vert H \vert m \rangle (G^k_m + (G^k_m)^\dagger),
\end{eqnarray}
where $G^k_m$ are defined through the mapping $\outerprod{m+k}{m} \to G^k_m$ (see \cite{rethinasamy2024neutronnucleus} for further details).

\begin{figure}
\centering
\includegraphics[width=0.82\columnwidth]{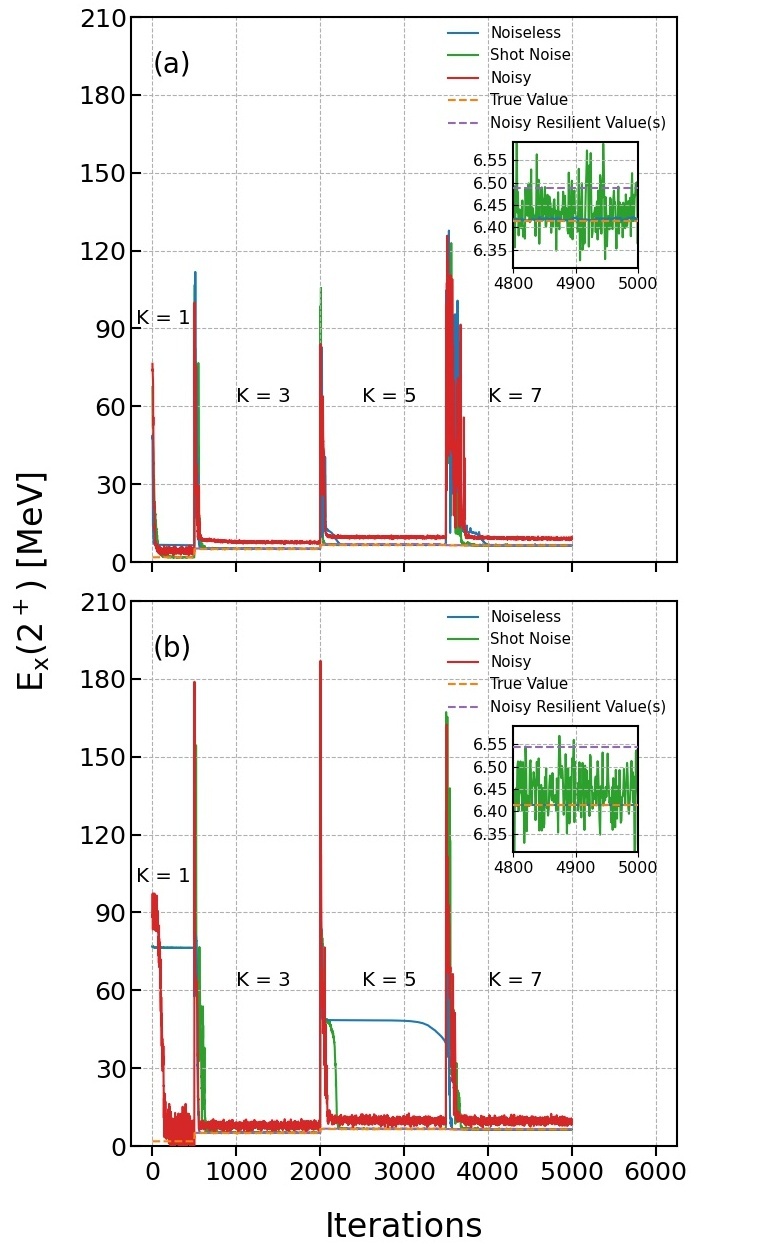}
\caption{\label{fig:grayvsOH}Quantum simulations using the (a) Gray ($n$ = 3 qubits) and (b) one-hot ($n$ = 8 qubits) encodings for the energy of the lowest $2^+$ state (relative to the theoretical energy of the $0^+$ state for each order $K$) of \textsuperscript{12}C with $N = 8$ basis states, starting with $K = 1$, as compared to the final theoretical energy $E_{\text{th}}^{K = 7}=6.414$ MeV, labeled as ``True Value”. Shown are:  noiseless (blue), shot noise (green), and noisy (red) simulations (see text for details). The inset plot is the last 200 iterations. }
\end{figure}

\begin{figure*}
\centering
\includegraphics[width=0.75\textwidth]{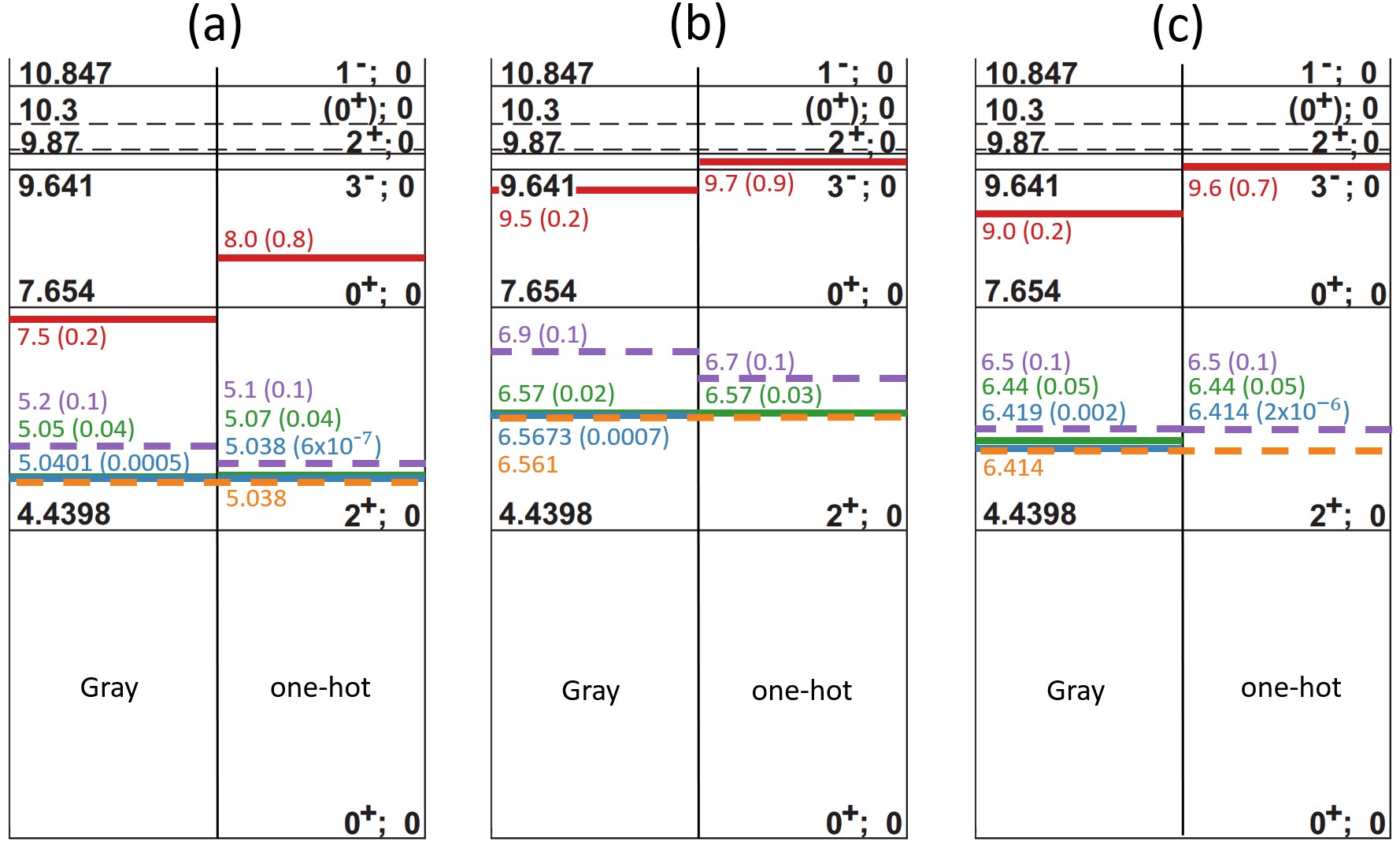}
\caption{\label{fig:grayvsOH2}Simulation energies of the lowest $2^+$ state (relative to the theoretical energy of the $0^+$ state for each order $K$) of \textsuperscript{12}C using the Gray ($n$ = 3 qubits) and one-hot ($n$ = 8 qubits) encodings  with $N = 8$  and with (a) $K = 3$, (b) $K = 5$, and (c) $K = 7$, reported as the mean and standard deviation of the last 100 iterations. The error in the noise-resilient energies are estimated from the standard deviation of 100 identical simulations. The color coding is identical to that in Fig.~\ref{fig:grayvsOH}. The experimental data (black lines) is labeled by the spin, parity, and energy (in MeV) of each state. 
}
\end{figure*}

We use a parameterized ansatz of the wavefunction, i.e., $\vert \psi(\boldsymbol{\theta}) \rangle=U(\boldsymbol{\theta})|0\rangle$. The quantum circuit to generate the entangled wavefunction for both encoding schemes can be found in \cite{rethinasamy2024neutronnucleus} and are shown in Figs. \ref{fig:Gray circuit} and \ref{fig:OH circuit}. We use the variational quantum eigensolver to minimize the energy of the system against the parameter $\boldsymbol{\theta}$. We diagonalize the full Hamiltonian on the classical computer and obtain the lowest energy of the $0^+$ and $2^+$ states, and report the  $2^+$ excitation energy relative to the ground-state energy (referred to as the ``true value", $E_{\rm th}^{K}$). This gives a lower bound on the ground-state energy.
As done in \cite{rethinasamy2024neutronnucleus}, we take advantage of a perturbative-like behavior of the nuclear Hamiltonian, namely, the lowest eigenvalues become slightly more accurate with increasing bandwidth $2K+1$.
Hence, we start with a lower value of $K$, where the Hamiltonian is reduced and needs fewer iterations to converge. The output-correlated wavefunction is used to initialize the simulation of a higher $K$ number, where more correlations are involved. This warm-start initialization is inspired by perturbation theory.

\section{Quantum simulations for\\ the $^{12}$C rotational band}

We provide quantum simulations of the lowest $2^+$ excitation energy, $E_{\rm x}(2^+)$, of $^{12}$C in an $N=8$ basis set (Fig.~\ref{fig:grayvsOH}). The Gray encoding scheme uses 3 qubits and the one-hot encoding scheme uses 8 qubits. 

We perform the following simulations. The exact diagonalization provides the ``true value" 
and is used to validate the quantum simulation outcomes. The noiseless simulations use a perfect noiseless simulator to provide ideal behavior. The shot-noise simulations use $1000$ shots. This is the noise from the final measurements, which cannot be avoided. These simulations provide a good  performance estimate for the far-term fault-tolerant regime. 

The noisy simulations use noise models from existing quantum devices. Specifically, in this study, the cost function estimates use a fake IBMQ backend {\tt ibm\_manila}. This provides a good performance estimate for the near-term NISQ regime. In the present study, these simulations are coupled with the noise-resilient technique, which was first introduced in ~\cite{Sharma_2020}. Namely, we use the final parameters $\boldsymbol{\theta}$ from the noisy simulation and calculate the expectation value of the Hamiltonian on a classical machine, which we refer to as the noise-resilient (NR) value.

For both encoding schemes, we start with a tridiagonal Hamiltonian ($K=1$), that is, we set all other matrix elements to zero, and gradually increase the bandwidth $2K+1$. As expected, the noiseless simulation in both cases gradually converges to the true value as $K$ increases (see the blue curve in Fig.~\ref{fig:grayvsOH}). We note that for $K=5$ the noiseless simulation with the one-hot encoding does not converge to the ``true value" within $1000$ iterations, likely because larger $K$ values may require larger number of iterations.

For both encodings, the shot-noise simulation achieves good agreement with the ``true value" (see the green curve in Fig.~\ref{fig:grayvsOH}), implying that increasing the number of qubits may not be inefficient in the fault-tolerant regime. However, the noisy simulation with the Gray encoding clearly yields results with less noise, as compared to the one-hot simulations (see the red curve in Fig.~\ref{fig:grayvsOH}). Based on this, we find that the noise-resilient estimate reproduces the ``true value". This implies that the algorithm exhibits noise resilience, that is, training is still possible in a noisy scenario. Therefore, even in the presence of noise, we expect that the current algorithm will obtain acceptable outcomes on near-term NISQ quantum devices.

Finally, we investigate how the simulation outcomes vary with increasing $K$ (Fig.~\ref{fig:grayvsOH2}), and whether one can utilize smaller values of the Hamiltonian matrix bandwidth for $^{12}$C, to take the full advantage of the Gray code. We note that \cite{rethinasamy2024neutronnucleus} has shown that simulations on actual quantum devices with the Gray encoding should outperform the one-hot encoding for band-diagonal Hamiltonians having a bandwidth up to $N$ (or equally, $K<N/2$) when commutativity schemes are employed to reduce the number of measurements. Indeed, as shown in Fig.~\ref{fig:grayvsOH2}a, the $K=3$ case already provides a reasonable excitation energy, including the shot-noise simulations and the final NR estimate. These estimates agree reasonably well with the experimental data and lie closer to the first $2^+$ excited state than to the second $2^+$ excited state at 9.87 MeV.
We note that the simulation energies of the lowest $2^+$ state are reported as the mean and standard deviation of the last 100 iterations. Consistently, the noisy simulations using the one-hot encoding have larger errors compared to the ones using the Gray code. The error in the noise-resilient energies are estimated from the standard deviation of 100 identical simulations, using the last outcome from the noisy simulation in each case, and does not propagate the errors we report for the noisy simulations. With the latter, the one-hot encoding is expected to yield larger errors for the NR values as well. 

\section{Conclusions}
In summary, we show that the noisy simulations that use the Gray encoding, coupled with the noise-resilient technique and the symmetry-adapted basis, reproduce reasonably well the lowest $2^+$ state of the ground-state rotational band of $^{12}$C, a twelve-particle quantum mechanical system. The noise-resilient method appears to be a reliable technique to improve the performance of near-term NISQ devices. The Gray encoding uses fewer qubits than the one-hot encoding, while showing an advantage in terms of stability and error, which is expected to improve with larger number of basis states, $N$. This holds promise for achieving acceptable solutions of problems in much larger model spaces and for heavier nuclei.

\bibliography{ORNLabstract_refs,lsu_latest}

\end{document}